\documentclass[aps,prl,showpacs,twocolumn,floats]{revtex4}
\usepackage{amssymb}

\usepackage{graphicx,psfig}
\usepackage{dcolumn}
\usepackage{bm}

\input{tcilatex}

\begin{document}

\title{Transitions at avoided level crossing with interaction and
disorder}
\author{D. A. Garanin and R. Schilling}
\affiliation{Institut f\"{u}r Physik, Johannes
Gutenberg-Universit\"{a}t, D-55099 Mainz, Germany}
\date{\today}

\begin{abstract}
We investigate the influence of interaction between tunneling
particles and disorder on their avoided-level-crossing transitions
in the fast-sweep limit. Whereas the results confirm expectations
based on the mean-field arguments that
ferromagnetic/antiferromagnetic couplings suppress/enhance
transitions, we found large deviations from the mean-field
behavior for dipole-dipole interactions (DDI) in molecular magnets
Mn$_{12}$ and Fe$_{8}$. For ideal crystals of the needle,
spherical, and disc shapes DDI tends to enhance transitions. This
tendency is inverted for the needle shape in the presence of even
small disorder in the resonance fields of individual particles,
however.
\end{abstract}

\maketitle

Transitions at avoided level crossing or the Landau-Zener (LZ) effect \cite
{lan32,zen32} is a well known quantum phenomena, mainly in physics of atomic
and molecular collisions. In the time-dependent formulation the dynamics is
that of a two-level system described by a pseudospin $s=1/2$ whose
Schr\"{o}dinger equation (SE) can be solved exactly for the linear sweep of
the effective longitudinal magnetic field \cite{zen32}. As the SE for a spin
1/2 is mathematically equivalent to the classical dissipationless
Landau-Lifshitz equation (LLE), the effect can be envisioned classically as
a rotation of a magnetization vector.

Recently the LZ effect was observed in the solid-state world on crystals of
molecular magnets Fe$_{8}$ \cite{weretal00epl} (see Refs.\ \cite
{wer01thesis,sesgat03}\ for a recent review). This posed a new major problem
of the LZ effect in many-body systems with interactions. As tunneling of one
particle between the two states changes conditions for the others [in this
case mainly via dipole-dipole interactions (DDI)] one is confronted, in
general, with a SE for $N\gg 1$ coupled two-level systems that contains $%
2^{N}$ time-dependent coefficients.

The problem can be simplified if one applies the mean-field approximation
(MFA) that considers one particle tunneling in the effective field that is a
sum of the externally sweeped field and the molecular field from other
particles that is determined self consistently. This is a model of the \emph{%
nonlinear} LZ effect that was applied to tunneling of the Bose-Einstein
condensate \cite{zobgar00,wuniu00}. Again this problem can be reformulated
in terms of a classical nonlinear LLE. The MFA solution shows that
ferromagnetic interactions suppress transitions while antiferromagnetic
interactions enhance them. Corrections to the MFA for a simplified
``spin-bag'' interaction model of $N$ pseudospins coupled to all others with
the same coupling $J$ \cite{hamraemiysai00,gar03prb} were studied in Ref.\
\cite{garsch03prb}. However there are no rigorous results for competing
interactions such as the DDI, while existing theories use postulated rate
equations (see, e.g., Ref.\ \cite{liuetal02prb}).

The aim of this Letter is to develop, for arbitrary interactions, an
expansion of $P$ for fast sweep rates $v$ where $P$ is close to 1. Whereas
the first term $\sim 1/v$ in $1-P$ is in most cases insensitive to the
interaction (that is why it was possible to extract the correct value of the
ground-state tunnel splitting $\Delta $ in Fe$_{8}$ \cite{weretal00epl}),
the next term $\sim 1/v^{2}$ does depend on the interaction and it shows
whether transitions are enhanced or suppressed. We take into account
inhomogeneities of individual resonances as they can strongly reduce the
effect of interaction. Then we apply our result to the DDI and show that its
effect differs considerably from the mean-field prediction.

We consider the transverse-field Ising model
\begin{equation}
\widehat{H}=-\frac{1}{2}\sum_{i}\left\{ \left[ H_{z}(t)-V_{i}\right] \sigma
_{iz}+\Delta \sigma _{ix}\right\} -\frac{1}{2}\sum_{i,j}J_{ij}\sigma
_{iz}\sigma _{jz},  \label{Ham}
\end{equation}
where $\mathbf{\sigma }_{i}$ are Pauli matrices, $H_{z}(t)=vt$ is the
time-linear sweep field, $V_{i}$ is the local shift of the resonance field, $%
\Delta $ is the splitting of adiabatic energy levels for $J_{ij}=0.$ The
initial state of our model is all pseudospins down. For $J_{ij}=0$ the well
known solution \cite{lan32,zen32} for the final-state probability for a spin
to remain in the initial state is
\begin{equation}
P\equiv P(\infty )=e^{-\varepsilon },\qquad \varepsilon \equiv \frac{\pi
\Delta ^{2}}{2\hbar v},  \label{PLZDef}
\end{equation}
whereas $P(t)$ can be expressed via hypergeometric functions. The fast-sweep
expansion of Eq.\ (\ref{PLZDef}) is $P\cong 1-\varepsilon +\varepsilon
^{2}/2-\ldots $ We will see that interaction $J_{ij}$ modifies the term $%
\varepsilon ^{2}.$ At order $\varepsilon ^{2}$ it is sufficient to take into
account maximally two spin flips out of the initial state and to write the
wave function in the form
\begin{eqnarray}
\Psi (t) &=&c_{0}(t)\left| \downarrow \downarrow \ldots \downarrow
\right\rangle +\sum_{i}c_{i}(t)\sigma _{i+}\left| \downarrow \downarrow
\ldots \downarrow \right\rangle  \nonumber \\
&&+\frac{1}{2!}\sum_{ij}c_{ij}(t)\sigma _{i+}\sigma _{j+}\left| \downarrow
\downarrow \ldots \downarrow \right\rangle +\ldots  \label{PsiDef1}
\end{eqnarray}
The one-particle staying probablity averaged over the sample is
\begin{equation}
P=1-\frac{1}{N}\sum_{i}\left| c_{i}\right| ^{2}-\frac{1}{N}\sum_{ij}\left|
c_{ij}\right| ^{2}-\ldots  \label{PDef}
\end{equation}
The Schr\"{o}dinger equation reads
\begin{eqnarray}
i\hbar \dot{c}_{0} &=&0\times c_{0}-\frac{\Delta }{2}\sum_{i}c_{i}  \nonumber
\\
i\hbar \dot{c}_{i} &=&E_{i}(t)c_{i}-\frac{\Delta }{2}c_{0}-\frac{\Delta }{2}%
\sum_{j}c_{ij}  \nonumber \\
i\hbar \dot{c}_{ij} &=&E_{ij}(t)c_{ij}-\frac{\Delta }{2}\left(
c_{i}+c_{j}\right) -\frac{\Delta }{2}\sum_{l}c_{ijl,}  \label{SchrEq}
\end{eqnarray}
etc. Here $E$ are the eigenvalues of the Hamiltonian with $\Delta =0$ and
the ground-state energy $E_{0}(t)$ subtracted and
\begin{eqnarray}
E_{i}(t) &=&-H_{z}(t)+\tilde{V}_{i},\qquad \tilde{V}_{i}\equiv
V_{i}+2\sum_{j}J_{ij}  \nonumber \\
E_{ij}(t) &=&-2H_{z}(t)+\tilde{V}_{i}+\tilde{V}_{j}-4J_{ij}.  \label{EDef}
\end{eqnarray}

For the fast linear sweep the solution of Eqs.\ (\ref{SchrEq})\ is a series
in the integer and half-integer powers of $\varepsilon $ that can be solved
by iterations starting from $c_{0}(t)=1$. It is sufficient to retain $c_{0},$
$c_{i},$ and $c_{ij}$ while $c_{ijl}$ can be dropped. Inserting found
coefficients into Eq.\ (\ref{PDef}) and calculating double and triple time
integrals yields the final result
\begin{equation}
P\cong 1-\varepsilon +\varepsilon ^{2}/2+\varepsilon ^{2}I_{0},\qquad I_{0}=%
\frac{1}{N}\sum_{ij}I_{ij}  \label{Peps2Fin}
\end{equation}
where
\begin{equation}
I_{ij}=A_{ij}+\cos \left( 2\pi \gamma _{ij}^{(0)}\beta _{ij}\right)
B_{ij}+\sin \left( 2\pi \gamma _{ij}^{(0)}\beta _{ij}\right) C_{ij},
\label{IDef}
\end{equation}
$I_{ij}=I_{ji},$ and $A_{ij},$ $B_{ij},$ $C_{ij}$ \ are defined by
\begin{eqnarray}
A_{ij} &=&\frac{1}{2}-\frac{1}{4}\left[ \frac{1}{2}-C\left( \gamma
_{ij}\right) \right] ^{2}-\frac{1}{4}\left[ \frac{1}{2}-S\left( \gamma
_{ij}\right) \right] ^{2}  \nonumber \\
&&-\frac{1}{4}\left[ \frac{1}{2}-C\left( \gamma _{ji}\right) \right] ^{2}-%
\frac{1}{4}\left[ \frac{1}{2}-S\left( \gamma _{ji}\right) \right] ^{2}.
\label{AijRes0}
\end{eqnarray}
\begin{eqnarray}
B_{ij} &=&-\frac{1}{2}\left[ \frac{1}{2}-C\left( \gamma _{ij}\right) \right] %
\left[ \frac{1}{2}-C\left( \gamma _{ji}\right) \right]   \nonumber \\
&&-\frac{1}{2}\left[ \frac{1}{2}-S\left( \gamma _{ij}\right) \right] \left[
\frac{1}{2}-S\left( \gamma _{ji}\right) \right] .  \label{BijRes0}
\end{eqnarray}
\begin{eqnarray}
C_{ij} &=&\frac{1}{2}\left[ \frac{1}{2}-C\left( \gamma _{ij}\right) \right] %
\left[ \frac{1}{2}-S\left( \gamma _{ji}\right) \right]   \nonumber \\
&&-\frac{1}{2}\left[ \frac{1}{2}-S\left( \gamma _{ij}\right) \right] \left[
\frac{1}{2}-C\left( \gamma _{ji}\right) \right] .
\end{eqnarray}
Here $C(x)$ and $S(x)$ are Fresnel integrals and
\begin{eqnarray}
\gamma _{ij} &\equiv &\alpha _{i}-\alpha _{j}+\beta _{ij},\qquad \gamma
_{ij}^{(0)}\equiv \alpha _{i}-\alpha _{j}  \nonumber \\
\alpha _{i} &\equiv &\frac{\tilde{V}_{i}}{\sqrt{2\pi \hbar v}},\qquad \beta
_{ij}\equiv \frac{4J_{ij}}{\sqrt{2\pi \hbar v}}=\frac{4J_{ij}}{\pi \Delta }%
\varepsilon ^{1/2}.  \label{gammaijDef1}
\end{eqnarray}

Eqs.\ (\ref{Peps2Fin})--(\ref{gammaijDef1}) is our main result that is valid
for arbitrary interactions and resonance shifts. Note that it has a pair
structure and thus it can be verified against the direct numerical solution
for the model of two coupled particles. Analytical form makes its
application practically possible; Triple time integrals that arise at the
intermediate stage cannot be computed numerically with a reasonable
precision within a reasonable time. In the homogeneous case $\gamma
_{ij}^{(0)}=0$ and $I_{ij}$ simplifies to
\begin{eqnarray}
I_{ij} &=&F\left( \beta _{ij}\right)   \nonumber \\
&=&C\left( \beta _{ij}\right) \left[ 1-C\left( \beta _{ij}\right) \right]
+S\left( \beta _{ij}\right) \left[ 1-S\left( \beta _{ij}\right) \right] .
\label{IijHomo}
\end{eqnarray}
The limiting forms of $F(\beta _{ij})$ are
\begin{equation}
F(\beta _{ij})\cong \left\{
\begin{array}{ll}
-\frac{3}{2}-\frac{2\sqrt{2}}{\pi \beta _{ij}}\cos \left( \frac{\pi }{2}%
\beta _{ij}^{2}+\frac{\pi }{4}\right) , & -\beta _{ij}\gg 1 \\
\beta _{ij}-\beta _{ij}^{2}, & |\beta _{ij}|\ll 1 \\
\frac{1}{2}-\frac{1}{\left( \pi \beta _{ij}\right) ^{2}}, & \beta _{ij}\gg 1.
\end{array}
\right.   \label{FLims}
\end{equation}
For the weak interaction Eq.\ (\ref{Peps2Fin}) then yields
\begin{equation}
P\cong 1-\varepsilon +\frac{\varepsilon ^{2}}{2}+\frac{4J_{0}}{\pi \Delta }%
\varepsilon ^{5/2},  \label{PJweak}
\end{equation}
a generalization of Eq.\ (26) of Ref.\ \cite{gar03prb} for the arbitrary
form of  $J_{ij}.$ Note that Eq.\ (\ref{PJweak}) is essentially a MFA result
as it only depends on the zero Fourier component $J_{0}$ of the coupling $%
J_{ij}.$ In contrast to thermodynamic systems, here the applicability of the
MFA is not controlled by the interaction radius alone. For the
nearest-neighbor interaction with $z$ neighbors the relative correction to
the last term of Eq.\ (\ref{PJweak}) is $-\left[ 4J_{0}/(\pi \Delta )\right]
\varepsilon ^{1/2}/z.$ Eq.\ (\ref{FLims}) shows that ferromagnetic
interactions, $J_{ij}>0,$ increase $P$ and thus suppress transitions,
whereas antiferromagnetic interactions facilitate transitions. The
saturation for strong ferro- and antiferromagnetic interactions in Eq.\ (\ref
{FLims}) corresponds to the case of well-separated resonances studied in
Sec. III of Ref.\ \cite{gar03prb}.

Let us proceed to the inhomogeneous case, $\gamma _{ij}^{(0)}\neq 0$. For $%
\left| \gamma _{ij}^{(0)}\right| -\max (|\beta _{ij}|,1)\gg 1,$ Eq.\ (\ref
{IDef}) yields
\begin{equation}
I_{ij}\cong \frac{\sqrt{2}}{\pi }\cos \left( \frac{\pi }{2}\gamma _{ji}^{2}+%
\frac{\pi }{4}\right) \frac{\beta _{ij}}{\gamma _{ij}^{(0)2}-\beta _{ij}^{2}}%
,  \label{IijStronggamma0}
\end{equation}
i.e., strong inhomogeneities reduce the effect of interactions as individual
resonances are well separated and flip of one particle does not bring
another particle past or before the resonance by the changing effective
field. For $|\beta _{ij}|\ll 1$ Eq.\ (\ref{IDef}) yields
\begin{eqnarray}
I_{ij} &\cong &\left[ \sin \left( \frac{\pi }{2}\gamma _{ij}^{(0)2}\right)
+\cos \left( \frac{\pi }{2}\gamma _{ij}^{(0)2}\right) \right] \beta _{ij}
\nonumber \\
&&+\left[ S\left( \gamma _{ij}^{(0)}\right) -C\left( \gamma
_{ij}^{(0)}\right) \right] \pi \gamma _{ij}^{(0)}\beta _{ij}.
\label{Iijweak}
\end{eqnarray}
For $\left| \gamma _{ij}^{(0)}\right| \ll 1$ this simplifies to $I_{ij}\cong
\beta _{ij}\left( 1-\pi \gamma _{ij}^{(0)2}/2\right) ,$ i.e., weak
inhomogeneities do not essentially suppress weak interactions. For $|\beta
_{ij}|-\left| \gamma _{ij}^{(0)}\right| \gg 1$ one obtains
\begin{equation}
I_{ij}\cong \left\{
\begin{array}{ll}
-1/2-\cos \left( 2\pi \gamma _{ij}^{(0)}\beta _{ij}\right) , & \beta _{ij}<0
\\
1/2, & \beta _{ij}>0.
\end{array}
\right.  \label{IijLargebeta}
\end{equation}

As the first line of Eq.\ (\ref{IijLargebeta}) corresponding to strong
antiferromagnetic coupling oscillates fast with $\gamma _{ij}^{(0)}$ the
limiting value $F=-3/2$ in Eq.\ (\ref{FLims}) is unstable with respect to
small inhomogeneities $V_{i}.$ The effect of inhomogeneities can be
accounted for by averaging Eq.\ (\ref{IDef}) over stochastic values of $%
\alpha _{i}$ with a normalized Gaussian distribution $\rho _{\alpha }(\alpha
)=\left( 2\pi \delta _{\alpha }^{2}\right) ^{-1/2}\exp \left[ -\alpha
^{2}/\left( 2\delta _{\alpha }^{2}\right) \right] $ and quadratic average $%
\left\langle \alpha _{i}^{2}\right\rangle =\delta _{\alpha }^{2}.$ The
distribution of $\gamma _{ij}^{(0)}=\alpha _{i}-\alpha _{j}$ is then given
by the same function with $\delta _{\alpha }^{2}\Rightarrow 2\delta _{\alpha
}^{2}.$ For $\delta _{\alpha }\ll 1$ one can just set $\gamma _{ij},\gamma
_{ji}\Rightarrow \beta _{ij}$ in $A_{ij},$ $B_{ij},$ and $C_{ij}$ in Eq.\ (%
\ref{IDef}) and use $\left\langle \cos \left( 2\pi \gamma _{ij}^{(0)}\beta
_{ij}\right) \right\rangle =e^{-\left( 2\pi \beta _{ij}\delta _{\alpha
}\right) ^{2}}.$ As this factor decays at $\beta _{ij}\sim 1/\delta _{\alpha
}\gg 1,$ one can further simplify the result and replace of $F(\beta _{ij})$
of Eq.\ (\ref{IijHomo}) by $\overline{F}(\beta _{ij})$ that satisfies $%
\overline{F}(\pm \infty )=\pm 1/2$:
\begin{eqnarray}
\overline{F}(\beta _{ij},\delta _{\alpha }) &=&F(\beta _{ij})+\delta F(\beta
_{ij},\delta _{\alpha }),\qquad \delta _{\alpha }\ll 1  \nonumber \\
\delta F(\beta _{ij},\delta _{\alpha }) &=&\theta (-\beta _{ij})\left[
1-e^{-\left( 2\pi \beta _{ij}\delta _{\alpha }\right) ^{2}}\right] .
\label{FAvrDef}
\end{eqnarray}
For $\beta _{ij}\ll 1$ from Eq.\ (\ref{Iijweak}) one obtains
\begin{eqnarray}
&&\overline{F}(\beta _{ij},\delta _{\alpha })\cong \beta _{ij}f\left( 2\pi
\delta _{\alpha }^{2}\right) ,\qquad f(x)\equiv \frac{x}{\sqrt{2(1+x^{2})}}
\nonumber \\
&&\times \left[ \frac{x+1}{\sqrt{\sqrt{1+x^{2}}+1}}-\frac{x-1}{\sqrt{\sqrt{%
1+x^{2}}-1}}\right] .  \label{FAvrSmallbeta}
\end{eqnarray}
Function $f(x)$ monotonically decreases and satisfies
\begin{equation}
f(x)\cong \left\{
\begin{array}{cc}
1-x/2, & x\ll 1 \\
1/\sqrt{2x}, & x\gg 1.
\end{array}
\right.  \label{flims}
\end{equation}

Let us now turn to the DDI between tunneling spins $\pm S$ of magnetic
molecules aligned along the $z$ axis:
\begin{equation}
J_{ij}=\frac{\left( g\mu _{B}S\right) ^{2}}{v_{0}}\phi _{ij},\qquad \phi
_{ij}=v_{0}\frac{3\cos ^{2}\theta _{ij}-1}{r_{ij}^{3}},  \label{JijDDI}
\end{equation}
where $v_{0}$ is the unit-cell volume, $r_{ij}$ is the distance between the
sites $i$ and $j$, and $\cos \theta _{ij}=r_{ij,z}/r_{ij}.$ We consider $S=10
$ molecular magnets Mn$_{12}$ having a tetragonal lattice with parameters $%
a=b=17.319$ \AA ,\ $c=12.388$ \AA\ ($c$ is the easy axis) and $v_{0}=abc=3716
$ \AA $^{3}$ and Fe$_{8}$ having a triclinic lattice with $a=10.52$ \AA\ ($a$
is the easy axis), $b=14.05$ \AA , $c=15.00$ \AA , $\alpha =89.9%
{{}^\circ}%
,$ $\beta =109.6%
{{}^\circ}%
,$ $\gamma =109.3%
{{}^\circ}%
$ and $v_{0}=abc\sin \alpha \sin \beta \sin \gamma =1971$ \AA $^{3}$ (see,
e.g., Ref.\ \cite{marchuaha01}) One can write $\beta _{ij}$ of Eq.\ (\ref
{gammaijDef1}) in the form
\begin{equation}
\beta _{ij}=\xi \phi _{ij},\qquad \xi \equiv \frac{4E_{D}}{\pi \Delta }%
\varepsilon ^{1/2}=\frac{4\left( g\mu _{B}S\right) ^{2}}{\pi \Delta v_{0}}%
\varepsilon ^{1/2}.  \label{xiDef}
\end{equation}
For Fe$_{8}$ $E_{D}=126.4$ mK and $\Delta \simeq 10^{-7}$ K, so that $%
4E_{D}/(\pi \Delta )\simeq 1.6\times 10^{6}$ and for not too fast sweep, $%
\varepsilon \sim 10^{-2},$ one has $\xi \sim 10^{5}.$ This is also an
estimation for the number of spins within the distance $r_{c}\equiv \left(
v_{0}\xi \right) ^{1/3}$ that strongly interact with a given spin.

\begin{figure}[t]
\unitlength1cm
\begin{picture}(11,6)
\centerline{\psfig{file=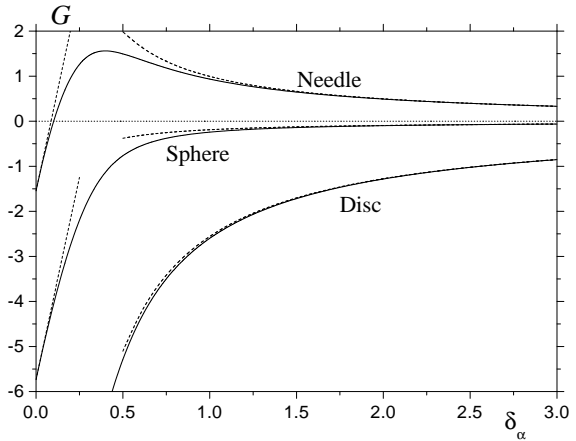,angle=-90,width=9cm}}
\end{picture}
\caption{ \label{Fig-LZQ-G}
$G$ of Eq.\ (\protect\ref{GDef}) for the sphere vs width of distribution of individual
resonances  $\delta_\alpha$. Dashed lines on the left and right are asymptotes of
Eqs.\ (\protect\ref{GSmalldeltaalpha})  and  (\protect\ref{GLargedeltaalpha}), respectively.
}
\end{figure}%
%

Consider a macroscopically large specimen of ellipsoidal form. According to
Eq.\ (\ref{IijLargebeta}) $I_{ij}$ does not diverge for $\beta
_{ij}\rightarrow \infty ,$ and for $\xi \gg 1$ one can replace the sum in
Eq.\ (\ref{Peps2Fin}) by an integral converging at $r_{ij}\sim r_{c},$ that
makes the result independent of the lattice structure$.$ Since at large
distances $I_{ij}\sim \beta _{ij}$ behaves as the DDI, the result depends on
the sample shape. For the model with random resonance shifts using $I_{ij}=%
\overline{F}(\beta _{ij},\delta _{\alpha })$ one obtains \
\begin{equation}
I_{0}\cong G\xi ,\qquad G=G^{(\mathrm{Sphere})}+\left( 1/3-n^{(z)}\right)
4\pi f\left( 2\pi \delta _{\alpha }^{2}\right) ,  \label{GDef}
\end{equation}
where $n^{(z)}=1/3,$ 0, and 1 for a sphere, needle and disc, respectively, $%
\ f(x)$ is that of Eq.\ (\ref{FAvrSmallbeta}), and
\begin{eqnarray}
&&G^{(\mathrm{Sphere})}=Kf\left( 2\pi \delta _{\alpha }^{2}\right) +\frac{%
8\pi }{9\sqrt{3}}\mathcal{P}\int_{-\infty }^{\infty }d\beta \frac{\overline{F%
}(\beta ,\delta _{\alpha })}{\beta ^{2}}  \nonumber \\
&&K\equiv -\frac{8\pi }{9}\left( 1-\frac{1}{\sqrt{3}}\ln \frac{\sqrt{3}+1}{%
\sqrt{3}-1}\right) =-0.66924.  \label{GSphereDef}
\end{eqnarray}
In general, $\overline{F}(x,\delta _{\alpha })$ is computed numerically from
Eq.\ (\ref{IDef}). For $\delta _{\alpha }\ll 1$ we use Eq.\ (\ref{FAvrDef})
that yields
\begin{equation}
G\cong -5.73432+16\left( \pi /3\right) ^{5/2}\left| \delta _{\alpha }\right|
+\left( 1/3-n^{(z)}\right) 4\pi .  \label{GSmalldeltaalpha}
\end{equation}
This result is non-analytical in $\delta _{\alpha }$ because random
inhomogeneities change the asymptotic behavior of $\overline{F}(\beta
,\delta _{\alpha })$ at $\beta \rightarrow -\infty .$ The large numerical
factor in Eq.\ (\ref{GSmalldeltaalpha}) makes $G^{(\mathrm{Sphere})}$ very
sensitive to $\delta _{\alpha }.$ For $\delta _{\alpha }\gg 1$ the
contribution of the integral term in Eq.\ (\ref{GSphereDef}) becomes
relatively small, and one obtains from Eq.\ (\ref{flims})
\begin{equation}
G\cong \frac{K+\left( 1/3-n^{(z)}\right) 4\pi }{2\sqrt{\pi }\delta _{\alpha }%
}.  \label{GLargedeltaalpha}
\end{equation}
Whereas $G<0$ for the sphere and disc, DDI acting predominantly
antiferromagnetically and enhancing LZ transitions, the result for the
needle in Eq.\ (\ref{GDef}) becomes positive already for $\delta _{\alpha
}\gtrsim 0.1$ (see Fig.\ \ref{Fig-LZQ-G}). As $\alpha _{i}\sim \left(
V_{i}/\Delta \right) \varepsilon ^{1/2}$ [see Eqs.\ (\ref{gammaijDef1}) and (%
\ref{PLZDef})], already resonance shifts $V_{i}$ of order $\Delta $ that can
stem from different sources yield $\alpha _{i}\sim \delta _{\alpha }\sim 0.1$
for sweep rates $\varepsilon \sim 10^{-2}.$

Let us compare our results with MFA result for $\delta _{\alpha }=0.$
\begin{eqnarray}
I_{0}^{(\mathrm{MFA})} &=&D_{zz}\xi ,\qquad D_{zz}\equiv \sum_{j}\phi _{ij}
\nonumber \\
D_{zz} &=&D_{zz}^{(\mathrm{Sphere})}+\left( 1/3-n^{(z)}\right) 4\pi .
\label{I0MFA}
\end{eqnarray}
Unlike $I_{0}$ in the limit $\xi \gg 1,$ the value of $I_{0}^{(\mathrm{MFA})}
$ depends on the lattice structure. For a simple cubic lattice $D_{zz}^{(%
\mathrm{Sphere})}=0$ and the result for $D_{zz}$ becomes purely macroscopic.
For tetragonal lattices $D_{zz}^{(\mathrm{Sphere})}>0$ if $a=b>c$ and $%
D_{zz}^{(\mathrm{Sphere})}<0$ if $a=b<c.$ Direct numerical calculation
yields $D_{zz}^{(\mathrm{Sphere})}=5.139$ for Mn$_{12}$ and $4.072$ for Fe$%
_{8}.$ Note that $E_{0}=-(1/2)D_{zz}E_{D}$ is the dipolar energy per site
for the ferromagnetic spin alignment. Our result $E_{0}=-4.131E_{D}$ for the
needle-shaped Fe$_{8}$ is in qualitative accord with $E_{0}=-4.10E_{D}$ of
Ref.\ \cite{marchuaha01}. One can see that $I_{0}^{(\mathrm{MFA})}>0$ for
the needle and sphere whereas $I_{0}^{(\mathrm{MFA})}<0$ for the disc, in
contradiction with our rigorous results above.

We have shown that the DDI generate huge corrections to the standard LZ
picture, $I_{0}\sim \xi \gg 1,$ because of its long-ranged character. Strong
nearest-neighbor interaction generate only moderate values of $I_{0},\ $%
e.g., $I_{0}$ $=z/2$ for the ferromagnetic coupling [see Eq.\ (\ref{FLims}%
)], and they cannot compete with the DDI. Our main result, Eq.\ (\ref
{Peps2Fin}), is applicable for $\varepsilon \left| I_{0}\right| \lesssim 1$
so that the term $\varepsilon ^{2}I_{0}$ is a correction to the leading term
$\varepsilon $ that defines the small transition probability $1-P$. The
theory breaks down in the slow-sweep range $\varepsilon \left| I_{0}\right|
\gtrsim 1,$ where the interaction strongly modifies the process.
Nevertheless Eqs.\ (\ref{Peps2Fin}), (\ref{xiDef}), and (\ref{GDef}) allow
to estimate the range of sweep rates where the single-particle description
of the LZ effect is valid and to see whether the interaction tends to
suppress or to enhance transitions. For the DDI the standard LZ effect can
be observed for
\begin{equation}
\varepsilon \lesssim \varepsilon _{c}=\left( \left| G\right| \frac{4E_{D}}{%
\pi \Delta }\right) ^{-2/3}  \label{epsiloncDef}
\end{equation}
that for Fe$_{8}$ results in a rather fast sweep rate $\varepsilon
_{c}\simeq 2.3\times 10^{-5}$ for a sphere without inhomogeneities ($G\simeq
-5.73$). This would preclude observation of a standard LZ effect in
experiments. However we have seen above that inhomogeneities of individual
resonances drastically reduce the effect of interaction and thus increase $%
\varepsilon _{c}.$

In the sweeping experiments \cite{weretal00epl} on the $\pm S$ transitions
in Fe$_{8}$ the sweep rate was $v=2Sg\mu _{B}dB/dt,$ so that with Eq.\ (\ref
{PLZDef}) one obtains $v_{c}=\pi \Delta ^{2}/(2\hbar \varepsilon _{c})$ and $%
\left( dB/dt\right) _{c}=\pi \Delta ^{2}/(4\hbar Sg\mu _{B}\varepsilon
_{c})\simeq 8\times 10^{-5}/\varepsilon _{c}\simeq 3$ T/s. However Fig.\ 2
of Ref.\ \cite{weretal00epl} shows that (i) standard LZ effect can even be
seen down to $\left( dB/dt\right) _{c}^{\exp }\sim 0.01$ T/s (i.e., $%
\varepsilon _{c}^{\mathrm{\exp }}\sim 10^{-2}$) and (ii) that $\Delta $ is
underestimated for $dB/dt\lesssim \left( dB/dt\right) _{c}^{\exp }.$ This
suggest that transitions are suppressed, i.e., ferromagnetic couplings are
dominating and $I_{0}>0.$ As in Ref.\ \cite{weretal00epl} a crystal of
rectangular shape ($l_{a}=80$ $\mu $m$,$ $l_{b}=50$ $\mu $m, $l_{c}=10$ $\mu
$m \cite{werpriv}) was used whose shape is closer to the needle than to the
sphere, the sign of the effect could be reconciled with our theory by
assuming even small random inhomogeneities$.$ On the other hand, for this
sample the inhomogeneities $\tilde{V}_{i}$ of Eq.\ (\ref{gammaijDef1}) are
of the order of the dipolar field itself that leads to much larger values of
$\varepsilon _{c},$ i.e., to slower sweep rates at which the interaction
precludes observing the standard LZ effect. These inhomogeneities are \emph{%
not random,} however, and the resonance shifts $\gamma _{ij}^{(0)}$ increase
with the distance between $i$ and $j,$ depending on the gradient of the
dipolar field. Dealing with this case would require using Eq.\ (\ref
{Peps2Fin}) with a complete solution for the inhomogeneous dipolar field in
the sample. While it can be done elsewhere, we recommend to perform
experiments on \ crystals of elliptic shape to avoid strong inhomogeneities
that suppress the effect of interaction and to see a more dramatic influence
of the DDI on the LZ transitions.

To conclude, quantum transitions in a system of many interacting two-level
particles is a tough problem that in general does not yield to familiar
approximate methods such as the mean-field approximation. We have calculated
rigorously the staying probability $P$ of the LZ effect in the fast-sweep
limit, $\varepsilon \ll 1,$ for general interactions. We have shown that
long-range interactions exceeding the level splitting $\Delta ,$ such as the
DDI, exert a profound influence upon the process. For spherical samples DDI
acts antiferromagnetically (contrary to the MFA predictions for Mn$_{12}$
and Fe$_{8})$ and enhances transitions. This should lead to overestimating
of the experimental value of $\Delta ,$ if the standard LZ formula, Eq.\ (%
\ref{PLZDef}), is used. To the contrary, for $\varepsilon \gtrsim 1$ one can
expect $P>e^{-\varepsilon }$, determined by those particles that are
prevented from tunneling by positive couplings$.$

\bibliographystyle{prsty}

\end{document}